\documentclass[manuscript]{aastex}

\slugcomment{Not to appear in Nonlearned J., 45.}

\shorttitle{Non-coplanar triple system V344 Lacertae}
\shortauthors{Liu et al.}

\begin{document}


\title{The contact binary V344 Lacertae: Is it a triple system?}


\author{Liu Liang\altaffilmark{1,2,3,4},
Qian Shengbang\altaffilmark{1,2,3,4}, Li Kai\altaffilmark{5}, He Jiajia\altaffilmark{1,2,3}, Li Linjia\altaffilmark{1,2,3}, Zhao Ergang\altaffilmark{1,2,3} and Li Xuzhi\altaffilmark{1,2,3,4}}


\altaffiltext{1}{Yunnan Observatories, Chinese Academy of Sciences, 396 Yangfangwang, Guandu District, Kunming, 650216, P. R. China (e-mail: LiuL@ynao.ac.cn)}
\altaffiltext{2}{Key Laboratory for the Structure and Evolution of Celestial Objects, Chinese Academy of Sciences, 396 Yangfangwang, Guandu District, Kunming, 650216, P. R. China}
\altaffiltext{3}{Center for Astronomical Mega-Science, Chinese Academy of Sciences, 20A Datun Road, Chaoyang District, Beijing, 100012, P. R. China}
\altaffiltext{4}{University of Chinese Academy of Sciences, Yuquan Road 19\#, Sijingshang Block, 100049 Beijing, China}
\altaffiltext{5}{Shandong Provincial Key Laboratory of Optical Astronomy and Solar-Terrestrial Environment, Institute of Space Sciences, Shandong University, Weihai, 264209, China}


\begin{abstract}
The $VRI$ passbands light curves of V344 Lac were presented and analyzed by using the latest version of the W-D code. The observed spectrum reveals that V344 Lac is not an A3 type but would be a later F type star according to the yielded temperature. The results of solution show that V344 Lac is an A-subtype contact binary, with a mediate photometric mass ratio of $0.387\pm0.003$ and a mediate contact factor of $44.6\pm3.0$\%. Based on the parallax given by Gaia, the parameters of the components are estimated as: $M_1=1.16\,\rm{M_{\odot}}$, $M_2=0.45\,\rm{M_{\odot}}$, $R_1=1.31\,\rm{R_{\odot}}$, $R_2=0.88\,\rm{R_{\odot}}$, $L_1=2.512\,\rm{L_{\odot}}$, $L_2=1.057\,\rm{L_{\odot}}$. The period investigation indicates that V344 Lac may have an eccentric orbital oscillation, with $P_3=12.4\pm0.5$\,yr, $A_3=0.0020\pm0.0002$\,d, and $e=0.38\pm0.16$. Analysis shows such oscillation would be caused by a magnetic activity which can be explained by the Applegate mechanism. Meanwhile, according to the value of $l_3$ and the estimated physical parameters of V344 Lac, the mass of the third companion may be $0.79M_{\odot}$. This third body could be a wide company.
\end{abstract}





\section{Introduction}
Contact binaries are binary stars where both components have already overflowed their Roche lobe so that they have formed a common radiative or convective envelope with very strong interactions \citep[e.g.][]{Qianetal2017}. Because of the contact in physical, they have the lowest angular momentum resulting from the tightest orbits among the same spectral type binaries, and they have very complex interactions. For example, the long-term cyclic behaviors of the orbital period would extend the life of contact binaries. A famous representative is the Thermal Relaxation Oscillations \citep[TRO,][]{Lucy1976,Flannery1976,RobertsonEggleton1977}, and the other possibility is the Common Convective Envelope Oscillations \citep[CCEO,][]{Liuetal2018}. Hence, there are still unsolved problems about the evolution of contact binaries.

That contact binaries are formed from closed detached binaries is widely accepted now. The migration of angular momentum between the central pair and the third companion may be a very important process for the formation of a contact binary \citep[e.g.,][]{Eggleton2012}. There are two reasons. First, contact binaries have been found in the young open cluster \citep[e.g. TX Cnc,][]{Liuetal2007,Zhangetal2009}, indicating that formational time-scale of contact binaries could be much shorter than the time-scale of their nuclear evolution. Second, third bodies have been found in many contact binary systems. \cite{PribullaRucinski2006} obtained $59\pm8\,\%$ of contact binaries which are brighter than the magnitude of 10 and are in the northern-sky have a third companion. \cite{Rucinskietal2007} have tried to prove that the contact binaries with $P < 1$ usually have an additional component by using of the adaptive optics technique. There are also some closed third bodies which were found by the combination of the $O-C$ method and the $l_3$ analysis from light curves (e.g., V345 Gem, \citealt{Yangetal2009}; MQ UMa, \citealt{Zhouetal2015}; V776 Cas, \citealt{Zhouetal2016}; V1005 Her, \citealt{Zhuetal2019}).

V344 Lacertae was discovered as an EW-type binary by \cite{MillerWachmann1973} in a visual variable survey. After its discovery, except for minima monitoring, the only one of the published light curve analysis was given by \cite{Pribullaetal2005a}, who yielded geometric elements of V344 Lacertae as $i = 77.5(2)$, $q = 0.24$, $f = 0.45(1)$. As \cite{MolikWolf2004} pointed out, V344 Lacertae is a special contact binary which is located in the blue envelope of the period-color diagram \citep[e.g.][]{Eggen1967}. However, after a spectroscopic observation, we found the spectral type of V344 Lac should be much later than A3-type. It may be not a contradictory with the period-color relationship.

In the present work, new photometric and spectroscopic data were presented in Section 2. Orbital period studies and light-curve analysis were given in Sections 3 and 4. Physical parameters of each component were estimated based on the Gaia parallax in Sections 5. In the last section, we interpreted the period variation.

\section{New photometric and spectroscopic data}
\begin{figure}
\begin{center}
\includegraphics[angle=0,scale=.8]{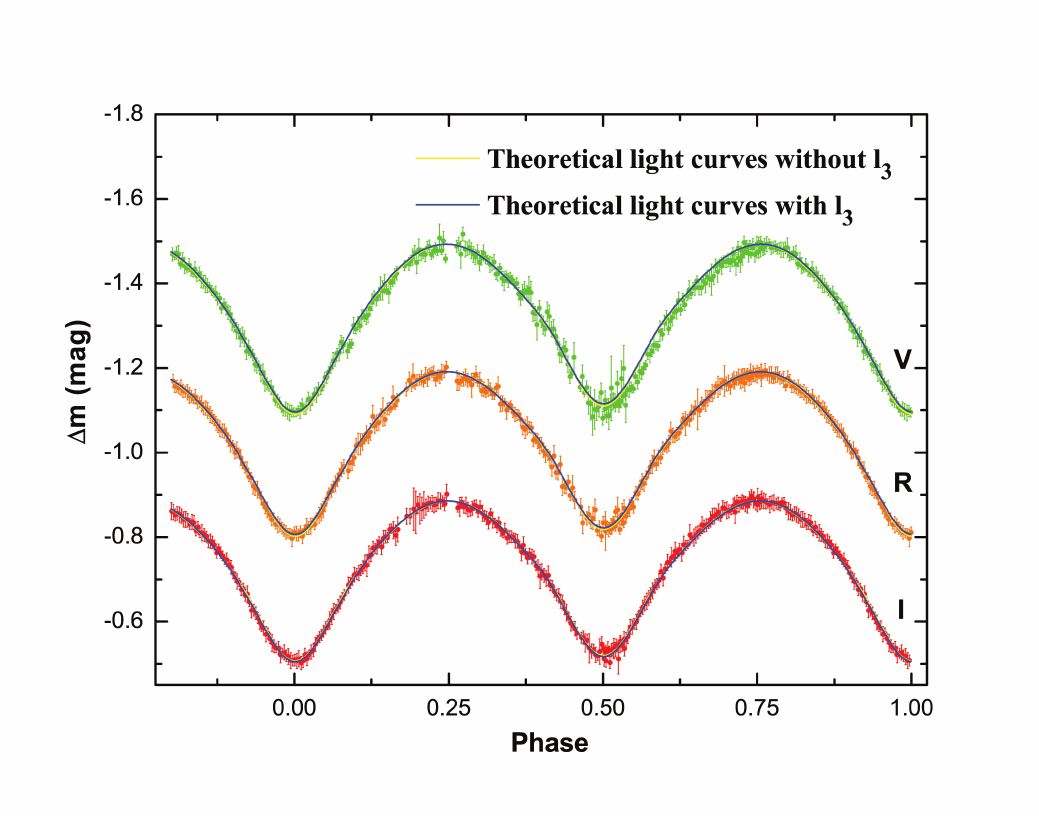}
\caption{Observed and theoretical light curves in $VRI$ passbands of V344 Lac.}
   \label{fig:org lc}
\end{center}
\end{figure}

V344 Lacertae, for the purpose of orbital period investigation, had been observed from 2009 to 2014 by the 1-m and 60-cm reflecting telescopes at the Yunnan Observatories (YNOs), and by the 85-cm reflecting telescope at the Xinglong Station, the National Astronomical Observatory (NAO). The 1-m and 60-cm telescopes were equipped with the same Andor DW436 CCD cameras, combining a standard Johnson-Cousins filter system. The 85-cm telescope was equipped with a primary-focus multi-color CCD photometer where a PI1024 BFT (Back-illuminated and Frame-Transfer) camera was used \citep{Zhouetal2009}, generating an effective field of view (FOV) as $16.'5 \times 16.'5$. And then on the nights of September 22, October 5 and 6, 2014, we obtained the three-color complete light curves (in $VRI$ filters) by the 1-m reflecting telescopes at the Weihai Observatory (WHO) of Shandong University, which was equipped a back-illuminated PIXIS 2048B CCD camera, with an FOV of 12'0 $\times$ 12'0 \citep{Huetal2014}. In all data reduction, 2MASS\,22184267$+$5157291 and 2MASS\,22181951$+$5156086 were chosen as comparison and check star, respectively. The aperture photometry package PHOT which is for measuring magnitudes for a list of stars in IRAF was used to reduce the observed images, after a standard flat-fielding correction process. Finally, the complete $VRI$ light curves and the others partly light curves for minima were obtained. The original data with the standard errors of the complete light curves can be found in our electronic table. Phases were calculated with the equation $2456923.06759+0.^{d}39224141 \times E$, where 2456923.06759 is the primary times of minimum obtained by the 1-m WHOT, $0.^{d}39224141$ is our corrected linear orbital period (section 3), and $E$ is the epoch number. The corresponding light curves are shown in Figure~\ref{fig:org lc}. Several times of minima obtained by us with the parabolic fitting method are listed in Table~\ref{tab: New minima of v344 Lac}.

\begin{figure}
\begin{center}
\includegraphics[angle=0,scale=.8]{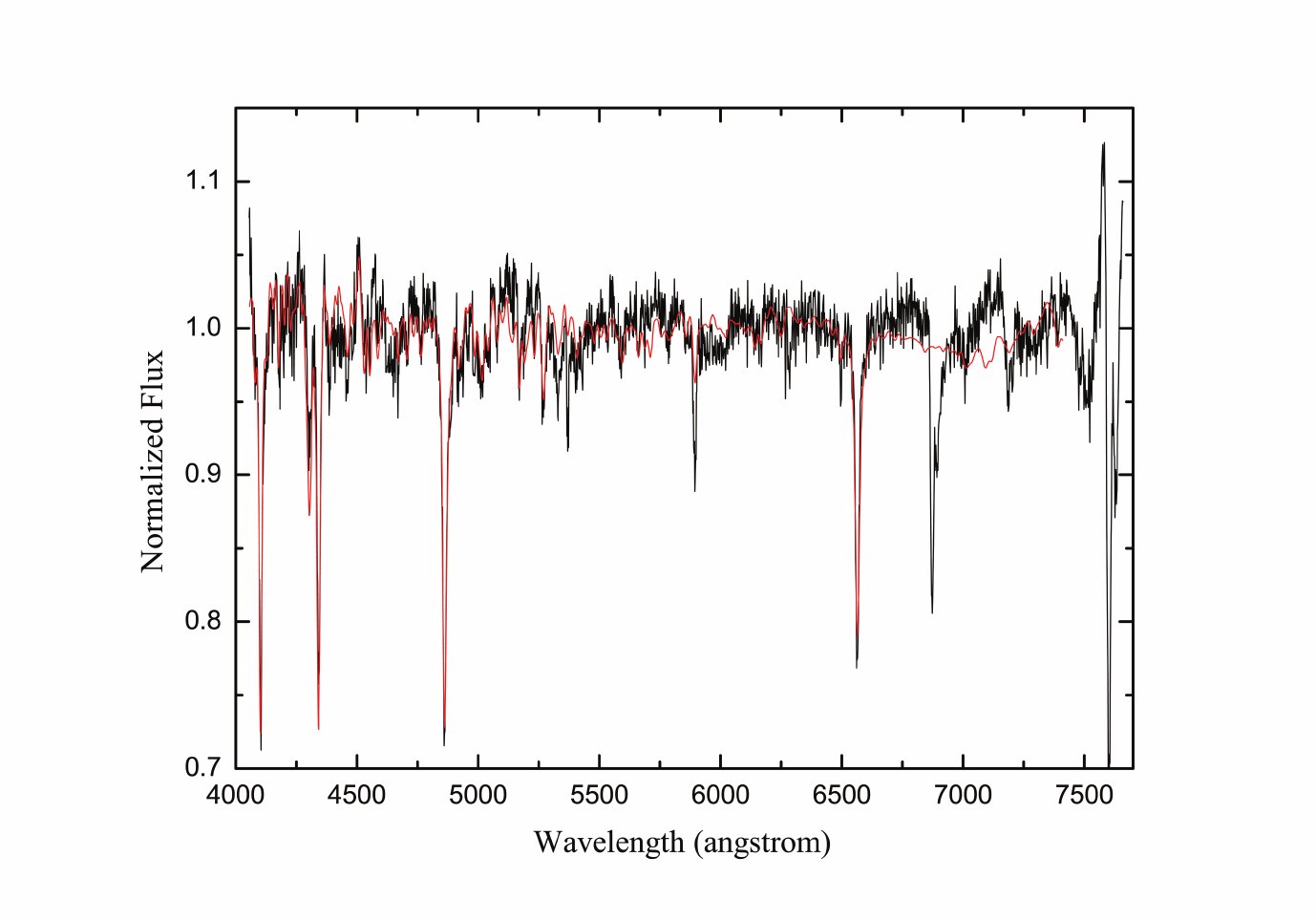}
\caption{Spectrum of V344 Lac. Black solid line is the observed spectrum, while the red solid line is the fitting one.}
    \label{fig:spec}
\end{center}
\end{figure}

To determine the accurate temperature of V344 Lac, spectroscopic observations were obtained with the Yunnan Faint Object Spectrograph and Camera (YFOSC) attached to the 2.4-m telescope in Lijiang Gaomeigu Station on December 13, 2018. Grism-14 of which the bandpass is from 320 to 750\,nm with a long-slit width of $2''.5$ was used \citep{Fanetal2015}. The exposure time was 600\,s. Bias subtraction and flat-field correction were made with standard IRAF procedures. The APEXTRACT package of the IRAF was used to extract the spectra. Wavelength calibration was performed using a He-Ne comparison lamp taken during the same observing run. University of Lyon Spectroscopic analysis Software (ULySS) package \citep{Kolevaetal2009,Wuetal2011} was used to analyze the spectra. The basic stellar atmospheric parameters were modeled as $T_{\rm{eff}}=6312\pm54$\,K and the metallicity [$Fe/H=-0.289\pm0.011$]. This value of temperature is similar to the result of Gaia \citep{GaiaCollaborationetal2018}. The corresponding spectral fitting is shown in the Figure~\ref{fig:spec}. The fitting of the spectral at the larger wavelength region is not good. It may be account for the effect of the cooler component/components, because the subsequent analysis will reveal that there may be a cool third company in the system.

\section{Orbital period analysis}
From 1948, the minima monitoring for V344 Lacertae has been over 70 years, which is helpful to study its long-term period variation. However, the data obtained by the visual or photograph method showed a very large scatter, so we discard those data in our analysis. Finally, 95 CCD or photoelectric times of minima of V344 Lac were collected and are listed in Table~\ref{tab: Minima of v344 Lac}, including 10 of ours. These data were used to correct the linear ephemeris starting with the formula of $2456923.06759(17)+0.^{d}39222768 \times E$. $T_0$ was one of our obtained minimum, and $P_0$ was obtained from \cite{SafarZejda2002} and \cite{Kholopovetal1985}. The corrected new linear ephemeris is,
\begin{eqnarray}
{\rm Min.\,I}&=&2456923.0683(\pm0.0005)+0.^{d}39224141(\pm0.00000005)\times{E}.
\end{eqnarray}
\noindent The $(O-C)$ values with respect to the linear ephemeris are listed in the sixth column of Table~\ref{tab: Minima of v344 Lac}. The corresponding $(O-C)$ diagram is displayed in Figure~\ref{fig:oc}. The general $(O-C)$ trend of V344 Lac shown in Figure~\ref{fig:oc}. In this figure, a possible cyclic variation is shown. We suppose that the variations in the $(O-C)$ diagrams are caused by the light-travel-time effect (LTTE) of a company orbiting in the binary system of elliptical orbit, then
\begin{eqnarray}
(O-C)&=&\Delta T + \Delta P \cdot E + \frac{\beta}{2}E^2\nonumber\\
&&+A_3[(1-e^2)\frac{\sin(\nu+\omega)}{1+e\cos\nu}+e\sin\omega].
\end{eqnarray}
\noindent This formula was given by \cite{Irwin1952,Irwin1959}, where $A_3=(a_{12}\sin i^{\prime})/c$. The Levenberg-Marquardt method was used in the non-linear fitting. The fitted parameters are listed in Table~\ref{tab: obital elements}, while Equation (2) is plotted as a solid line in the upper panel of Figure~\ref{fig:oc}. The final residuals of $(O - C)$ are tabulated in the seventh column of Table~\ref{tab: Minima of v344 Lac} and are shown in the lower panel of Figure~\ref{fig:oc}.

\section{Photometric solutions of V344 Lac}
\begin{figure}
\begin{center}
\includegraphics[angle=0,scale=.8]{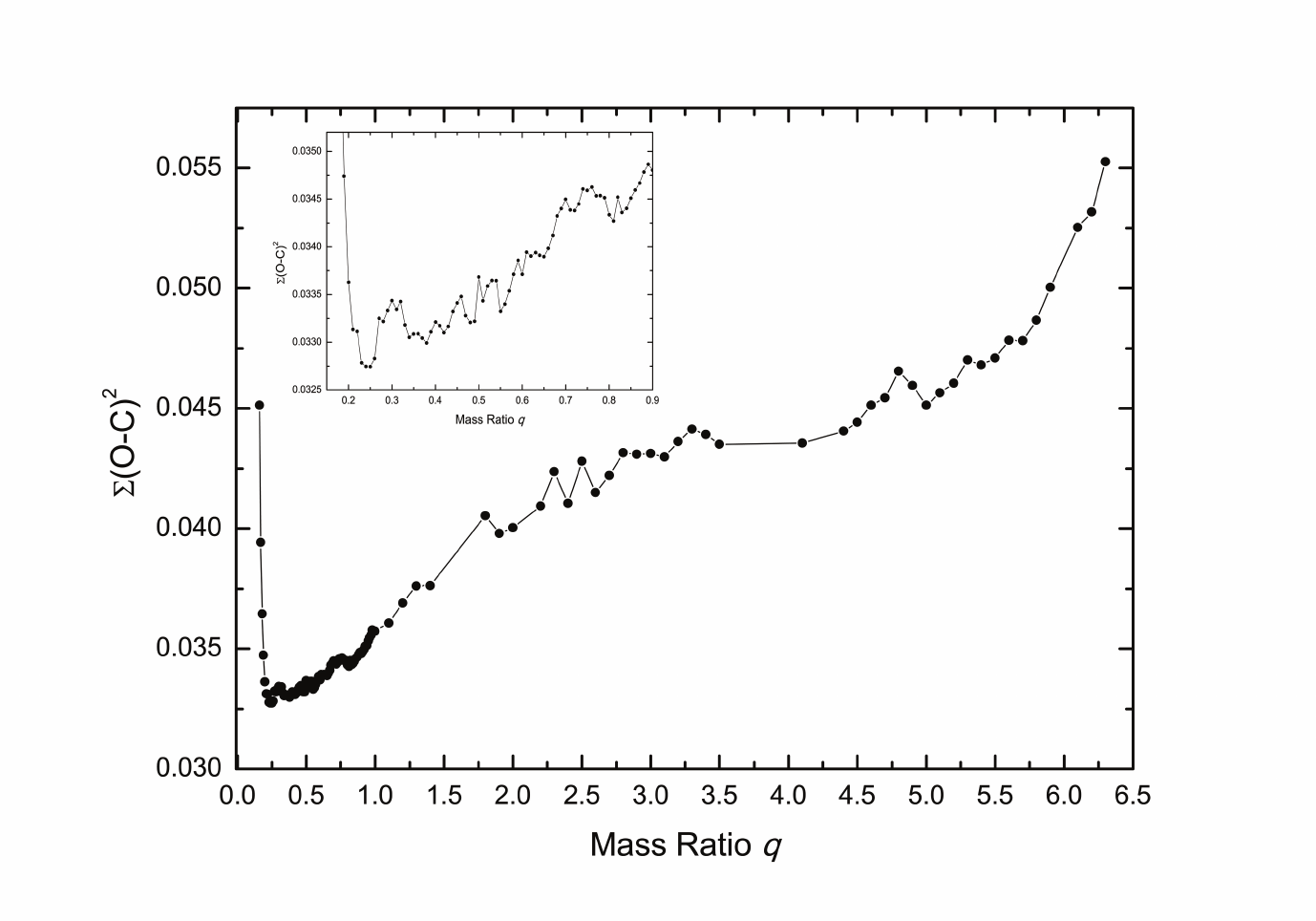}
\caption{The relationship between the mass ratio $q$ and the fitting residuals.}
    \label{fig:qs}
\end{center}
\end{figure}

To determine photometric elements and further to understand the geometrical structure and evolutionary state of V344 Lac, the 2015 Version of the W-D code \citep{WilsonDevinney1971, Wilson1979, Wilson1990, Wilson2008, Wilson2012, vanHammeWilson2007, Wilsonetal2010, WilsonvanHamme2014} was used to analyze the obtained light curves. The $q$-search method was employed to find an initial value of $q$. We fixed $q$ at a series of values of 0.1, 0.2, 0.3, etc., then fitted the light curves with the W-D code for each $q$ value, obtaining a series of corresponding fitting residuals, shown in Figure~\ref{fig:qs}, and selected the value of $q$ with minimal residuals as the input parameter. The mass ratio thus should be wide range from 0.2 to 0.6 under each model.

\begin{figure}
\begin{center}
\includegraphics[angle=0,scale=1]{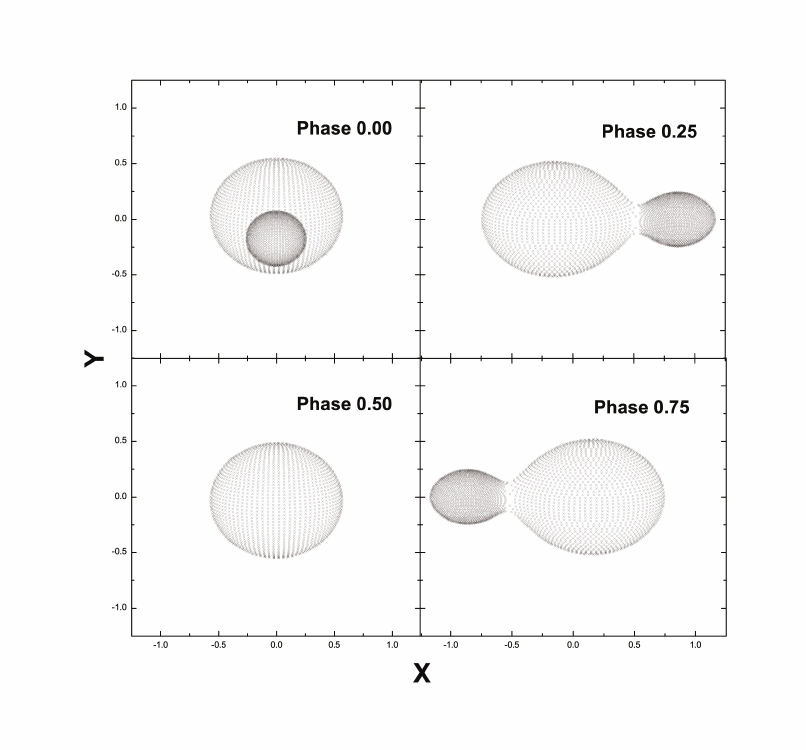}
\caption{Geometric structure of V344 Lac at phase 0.00, 0.25, 0.50 and 0.75.}
    \label{fig:str}
\end{center}
\end{figure}

During the solution, the bolometric albedo $A_1=A_2=0.5$ with the gravity-darkening coefficient of $g_1=g_2=0.32$ for convective equilibrium are used \citep{Lucy1967,Rucinski1969}, respectively. According to \cite{ClaretGimenez1990}, the square root limb-darkening coefficients were used. We adjusted the mass ratio $q$; the mean temperature of star\,2, $T_2$; the monochromatic luminosity of star\,1 and the dimensionless potential of star\,1. We started the DC program at model\,2. After thousands of iterations it converged to model\,3 ($\Omega_1=\Omega_2$, mode\,3 for contact configuration). Because the $(O-C)$ diagram presented some cyclic variation, we ran the program with the adjusted parameter of $l_3$ light. The photometric solutions are listed in Table~\ref{tab: Photometric solutions for V344 Lac} and the theoretical light curves computed with those photometric elements are plotted in Figure~\ref{fig:org lc}, while the geometric constructions of V344 Lac are shown in Figure~\ref{fig:str}.

\section{Estimation of the physical parameters with the Gaia parallax}
Physical parameters such as mass, radius and luminosity are very important information for a contact binary system. Hence it is necessary to estimate them. Here, we will introduce how to estimate the physical parameters without radial velocity curves.

We have known that the Parallax of V344 Lac is $1.6266\pm0.0363$\,mas \citep[Gaia DR2][]{GaiaCollaborationetal2018}. According to the formula of
\begin{equation}
(m-M)_V=10-5{\rm{lg}}\,{\rm{para}},
\end{equation}
we obtain a distance modulus of $8.944\pm0.049$\,mag. The V-band magnitude of the comparison star (2MASS 22184267+5157291) is 13.792 mag \citep[URAT1 Catalog,][]{Zachariasetal2015}. According to our V-band light curve (Figure~\ref{fig:org lc}), the magnitude of V344 Lac should be from 12.299 to 12.697 mags in the V-band. The composite V-band absolute magnitude therefore is about 3.355 mags. Adopted a bolometric correction (e.g., BCv $= -0.027$ mag for $T=6312$\,K, \citealt{WortheyLee2011}), the absolute bolometric magnitude of V344 Lac should be 3.328 mags.
After a bolometric correction, the bolometric absolute magnitude and the total luminosity are obtained with the relationships
\begin{equation}
M_{\rm{bol}}=M_V+\rm{BCv},
\end{equation}
\begin{equation}
L_T=L_1+L_2+L_3=10^{-0.4(M_{\rm{bol}}-4.83)}\cdot{\rm{L_{\odot}}}.
 \end{equation}
BCv is $-0.027$\,mag for the temperature of 6312\,K \citep{WortheyLee2011}. The total luminosity obtained in this calculation is independent of the photometric solution. According to the law of blackbody, we have
\begin{equation}
L_1+L_2=(\frac{T_1}{T_{\odot}})^4(Ar_1)^2+(\frac{T_2}{T_{\odot}})^4(Ar_2)^2,
\end{equation}
where $T_{\odot}$ is the effective temperature of the Sun (adopting 5770\,K); $A$ is the separation of the two components; $r_1$ and $r_2$ are the relative radii for star\,1 and star\,2, respectively ($r_i=(r_{i\rm{pole}}{\cdot}r_{i\rm{side}}{\cdot}r_{i\rm{back}})^{1/3}$). Combine Eq 5 and 6, we obtain $A$ as follows,
\begin{equation}
A=[10^{-0.4(M_{\rm{bol}}-4.83)}-L_3]^\frac{1}{2}[(\frac{T_1}{T_{\odot}})^4r_1^2+(\frac{T_2}{T_{\odot}})^4r_2^2]^{-\frac{1}{2}}.
\end{equation}
The contribution of $l_3$ are known from the photometric solution in Section 4. Then, by using of the Kepler's third law,
\begin{equation}
\frac{A^3}{P^2}=74.5(M_1+M_2),
\end{equation}
where $A$ is in unit of solar radius, while $P$ is in days, and $M_1, M_2$ are in solar mass. The total mass therefore can be determined. Because the mass ratio $q=M_2/M_1$ has been obtained, the individual mass of component is also known. Hence, the absolute physical parameters are estimated as follows, $M_1=1.16\,\rm{M_{\odot}}$, $M_2=0.45\,\rm{M_{\odot}}$, $R_1=1.31\,\rm{R_{\odot}}$, $R_2=0.88\,\rm{R_{\odot}}$, $L_1=2.512\,\rm{L_{\odot}}$, $L_2=1.057\,\rm{L_{\odot}}$.

Uncertainties of this estimation can be given by the error transfer formula,
\begin{equation}
\Delta M_{\rm{bol}}= \Delta m_V+ \frac{5}{{\rm ln}\,10}\frac{\Delta para}{para}+\Delta \rm{BCv},
\end{equation}
and
\begin{equation}
\frac{\Delta M}{M_{\rm{tot}}}=\frac{3}{2}\frac{\Delta L_T+\Delta L_3}{L_T-L_3}+\frac{6r_1^2T_1^3\Delta T_1+3T_1^4r_1\Delta r_1+6r_2^2T_2^3\Delta T_2+3T_2^4r_2\Delta r_2}{T_1^4r_1^2+T_2^4r_2^2}+\frac{2\Delta P}{P}.
\end{equation}
\noindent Having ignored the high-order-small-quantities, we have
\begin{equation}
\frac{\Delta M}{M_{\rm{tot}}} \simeq \frac{0.6{\rm ln}\,10\cdot L_T\Delta M_{\rm{bol}}+1.5\Delta L_3}{L_T-L_3}+\frac{6r_1^2T_1^3\Delta T_1 + 6r_2^2T_2^3\Delta T_2}{T_1^4r_1^2+T_2^4r_2^2}.
\end{equation}
\noindent In most situations, $\Delta T/T < 0.01$. Hence, for a safe estimation, $\Delta M/M_{\rm{tot}} \simeq 0.14$ when $\Delta M_{\rm{bol}} = 0.1$\,mag.

\section{Discussion and conclusions}
By the solution of the multiple color light curves, we found that V344 Lac is a moderate mass ratio ($0.387\pm0.003$), moderate contact binary ($44.6\pm3.0$\%). The more massive component is the hotter one. Hence it is an A-subtype contact binary system \citep{Binnendijk1970}. Combined with the parallax given by Gaia DR2, the physical parameters of each component are estimated.

\begin{figure}
\begin{center}
\includegraphics[angle=0,scale=.8]{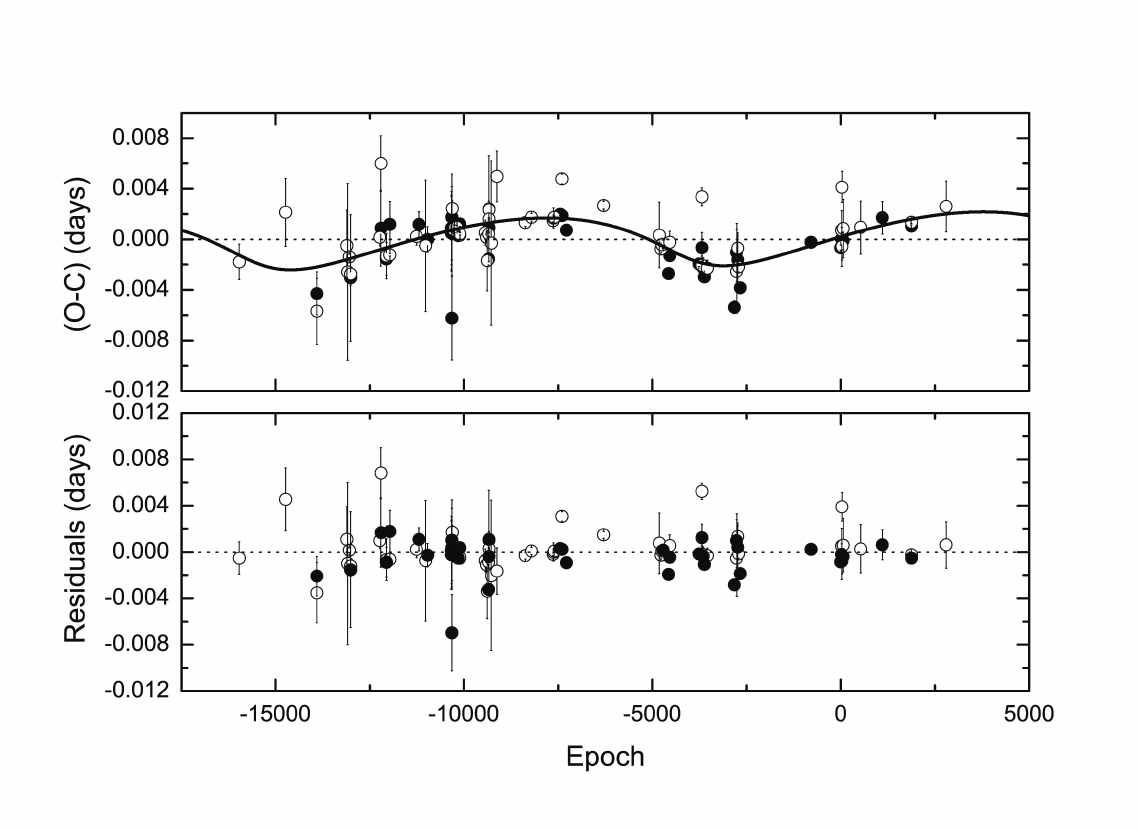}
\caption{The $(O-C)$ diagram of V344 Lac was formed by all CCD available measurements. The $(O-C)$ values were computed by using a newly determined linear ephemeris (Eq\,1). Full circles refer to the primary times of minima and empty ones to the secondary times of minima. Solid line represents the cyclic fitting (Eq\,2), while the dashed line is a zero line.}
    \label{fig:oc}
\end{center}
\end{figure}

\begin{figure}
\begin{center}
\includegraphics[angle=0,scale=.8]{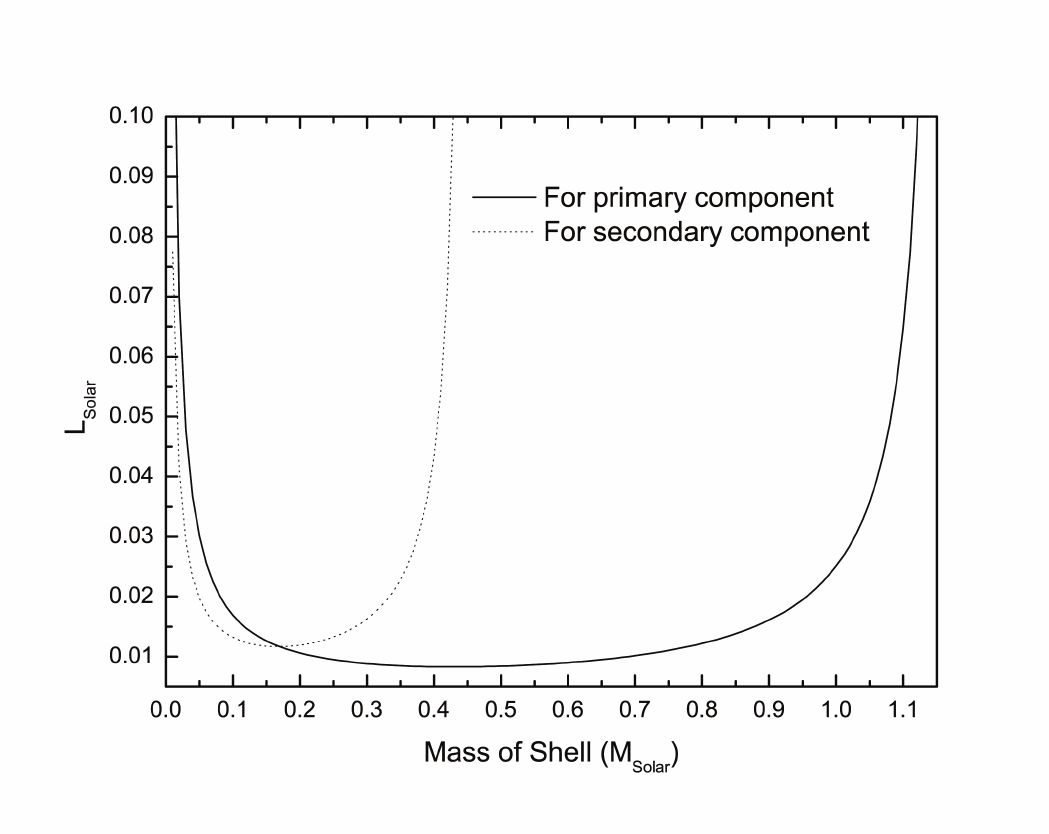}
\caption{Luminosity required to produce the cyclic oscillation in the $(O-C)$ diagram using the Applegate's mechanism. The solid line denotes the required luminosity for the primary component (more massive one), while the dashed line denotes that for the secondary. }
    \label{fig:applegate energy}
\end{center}
\end{figure}

Based on all available photoelectric and CCD eclipsing times, the possible period changes of V344 Lac were analyzed in Section 3. The orbital period was revised as 0.39224141 day by using the 95 timings being listed in Table~\ref{tab: Minima of v344 Lac}. The orbital period shows an eccentric cyclic variation (Figure~\ref{fig:oc}). Because both the primary and the secondary times of light minimum follow the same general trend of $(O-C)$ variation, it is excluded that the oscillation was caused by the apsidal motions. Cyclic period change can be caused by the mechanism of magnetic activity \citep[e.g.,][]{Applegate1992,Lanzaetal1998}. Figure~\ref{fig:applegate energy} shows the required luminosity to drive the Applegate mechanism. It is a very small fraction of the luminosity for both components. Hence the Applegate mechanism could be an explanation of the period oscillation. There was not an obvious cool spot in our observed light curves could be on account of the halcyon period according to location of time in the $(O-C)$ diagram. On the other hand, because of the solution of $l_3$, the LTTE due to the presence of a third body \citep[e.g.,][]{LiaoQian2010} is also discussed here. With the equation of
\begin{equation}
f(m)=\frac{(m_{3}\sin{i^{\prime}})^{3}} {(M_{1}+M_{2}+m_{3})^{2}},
\end{equation}
it is obtained that the minima mass (at $i^{\prime}=90^{\circ}$) of the third body is estimated to be $m_{3min} = 0.10\pm0.05\,M_{\odot}$, and that an orbital semimajor axis shorter than $6.40\pm0.61$ au. However, according to the solution and the estimated physical parameters of V344 Lac, $L_3$ could be $0.397\,\rm{L_{\odot}}$. The mass of a main-sequence star with such luminosity is about 0.79 solar mass, corresponding to a K0 type star \citep{Cox2000}. The mass of the third body will reach this value only while $i^{\prime}$ is very small.

Some well-studied contact binaries which have cyclic oscillation in their period are collected in Table~\ref{tab: cyclic variation of contact binaries}. The eleventh column denotes detection of cool spot in their light curves. In this table, we find that most of the oscillations are non-eccentric, and 12 of the 35 were shown $l_3$ in their photometric solutions. Cool spots and third light have been found in the same time in five samples. Especially, V776 Cas has been found a third light both by the spectroscopic method \citep{D'Angeloetal2006} and the photometric solution \citep{Zhouetal2016}. However, the third body of V776 Cas may be a wide company \citep[The separation is about 5.38 arcsec,][]{D'Angeloetal2006}. Hence, for V344 Lac, the $l_3$ in photometric solution may be on account of a wide K0 type third body, while the oscillation in orbital period should be due to a 12.4-years-magnetic-activity.

The third body in a contact binary system is a common phenomenon. Several studies suggest that a similarly large fraction of contact binaries have a third component \citep{PribullaRucinski2006,D'Angeloetal2006,Rucinskietal2007}; even the fraction is as large as 96\% \citep{Tokovininetal2006}. The probability of a third body in V344 Lac is therefore high.

The third body also would effect the formation of V344 Lac. \cite{FabryckyTremaine2007} investigated the orbital shrinkage by Kozai effect \citep{Kozai1962} and tidal friction, concluding that recently binary stars with orbital periods of 0.1$\sim$10 days are produced from binaries with much longer periods (10 $\sim10^5$ days) via the presented tertiaries. The same conclusion was obtained by \cite{Qianetal2018} via a statistic work for EA-type binaries. On the other hand, \cite{Qianetal2013a,Qianetal2013b,Qianetal2014b} had emphasized for many times that the relative close companion played an important role by exchanging angular momentum between the binary and the companion, which reduces angular momentum of central close binaries then forming a compact initial orbit. V344 Lac may have been formed in this way, too.

\acknowledgments
{
We are grateful to the anonymous referee who has given very useful suggestions to improve the paper. This work is partly supported by the Chinese Natural Science Foundation (No.\,11773066), the young academic and technology leaders project of Yunnan Province (No.\,2015HB098), and the Open Project Program of the Key Laboratory of Optical Astronomy, National Astronomical Observatories, Chinese Academy of Sciences. New observations of the systems were obtained with the 2.4-m telescope of Lijiang station of Yunnan Observatories, the 1-m telescope at the Weihai Observatory (WHO) of Shandong University, with the 85-cm telescope at Xinglong observation base, and with the 1-m and 60-cm telescope at Yunnan Observatories (YNOs). We would like to thank all the staff of the 2.4-m telescope of Lijiang station of Yunnan Observatories for their help and support during our observations. We also thank the persons who observed or reduced the data.
}

\begin{table}
\tiny
\caption{New times of light minima for V344 Lac.}
\label{tab: New minima of v344 Lac}
\begin{center}
\begin{tabular}{lllll}\hline
J.D. (Hel.) (d) & Error (d) & $Min.$ & Filter & Telescope\\
\hline
2455071.29617  & 0.00033  & s   &  $R$ & XL 85cm   \\
2455132.09126  & 0.00012  & s   &  $R$ & YNO 60cm  \\
2455492.16933  & 0.00080  & s   &  $R$ & YNO 60cm  \\
2455501.19018  & 0.00025  & s   &  $R$ & YNO 60cm  \\
2455535.11975  & 0.00063  & p   &  $R$ & YNO 60cm  \\
2455816.15762  & 0.00050  & s   &  $R$ & YNO 60cm  \\
2455876.17208  & 0.00047  & s   &  $R$ & YNO 1m    \\
2456923.06757  & 0.00017  & p   &  $V$ & WHO 1m    \\
2456923.06757  & 0.00013  & p   &  $R$ & WHO 1m    \\
2456923.06762  & 0.00016  & p   &  $I$ & WHO 1m    \\
2456936.20572  & 0.00108  & p   &  $V$ & WHO 1m    \\
2456936.20833  & 0.00118  & p   &  $R$ & WHO 1m    \\
2456936.20945  & 0.00103  & p   &  $I$ & WHO 1m    \\
2456936.99921  & 0.00107  & p   &  $V$ & WHO 1m    \\
2456936.99631  & 0.00160  & p   &  $R$ & WHO 1m    \\
2456936.99553  & 0.00105  & p   &  $I$ & WHO 1m    \\
\hline
\end{tabular}
\end{center}

Note: "XL 85cm" refers to the 85cm telescope at the Xinglong Station, the National Astronomical Observatory. "YNO 60cm" and "YNO 1m" refer to the 60cm and 1m telescopes at the Yunnan Observatories, respectively, while "WHO 1m" denotes the 1m telescope at the Weihai Observatory (WHO) of Shandong University.

\end{table}


\begin{deluxetable}{lllrrrrl}
\tabletypesize{\tiny}
\tablecaption{All available Times of light minima with errors for V344 Lac.}
\tablewidth{0pt}
\tablehead{
\colhead{JD (Hel.)} &  \colhead{Filter} & \colhead{Min.} & \colhead{Error (d)} &
\colhead{Epoch} & \colhead{$(O-C)$ (d)} & \colhead{Residuals (d)} & \colhead{Reference}\\}
\startdata
2450658.3828  &   $       $& s & $0.0014 $  &$-15971.5$  & $-0.00177  $ & $-0.00052  $& IBVS No. 4887          \\
2451142.4126  &   $       $& s & $0.0027 $  &$-14737.5$  & $0.00213   $ & $0.00453   $& IBVS No. 5263          \\
2451467.3782  &   $V      $& p & $0.0017 $  &$-13909  $  & $-0.00428  $ & $-0.00209  $& IBVS No. 5263          \\
2451467.5729  &   $V      $& s & $0.0026 $  &$-13908.5$  & $-0.00570  $ & $-0.00351  $& IBVS No. 5263          \\
2451782.5480  &   $       $& s & $0.0028 $  &$-13105.5$  & $-0.00049  $ & $0.00111   $& OEJV No. 0074          \\
2451786.4683  &   $       $& s & $0.0007 $  &$-13095.5$  & $-0.00257  $ & $-0.00098  $& IBVS No. 5484          \\
2451806.4738  &   $       $& s & $0.0002 $  &$-13044.5$  & $-0.00138  $ & $0.00017   $& IBVS No. 5484          \\
2451817.4552  &   $       $& s & $0.0009 $  &$-13016.5$  & $-0.00274  $ & $-0.00122  $& IBVS No. 5484          \\
2451817.651   &   $       $& p & $0.005  $  &$-13016  $  & $-0.00306  $ & $-0.00154  $& IBVS No. 5484          \\
2452123.4064  &   $       $& s & $0.0007 $  &$-12236.5$  & $0.00016   $ & $0.00099   $& IBVS No. 5484          \\
2452134.3950  &   $       $& s & $0.0022 $  &$-12208.5$  & $0.00600   $ & $0.00680   $& IBVS No. 5484          \\
2452134.586   &   $       $& p & $0.003  $  &$-12208  $  & $0.00088   $ & $0.00168   $& IBVS No. 5484          \\
2452194.4006  &   $       $& s & $0.0018 $  &$-12055.5$  & $-0.00133  $ & $-0.00067  $& IBVS No. 5484          \\
2452194.5965  &   $       $& p & $0.0013 $  &$-12055  $  & $-0.00155  $ & $-0.00089  $& IBVS No. 5484          \\
2452228.3320  &   $       $& p & $0.0018 $  &$-11969  $  & $0.00119   $ & $0.00178   $& IBVS No. 5484          \\
2452228.5257  &   $       $& s & $0.0001 $  &$-11968.5$  & $-0.00123  $ & $-0.00064  $& IBVS No. 5484          \\
2452505.4496  &   $       $& s & $0.0007 $  &$-11262.5$  & $0.00023   $ & $0.00021   $& IBVS No. 5484          \\
2452531.5346  &   $       $& p & $0.0010 $  &$-11196  $  & $0.00118   $ & $0.00110   $& IBVS No. 5438          \\
2452592.3311  &   $       $& p & $0.0010 $  &$-11041  $  & $0.00021   $ & $0.00015   $& IBVS No. 5438          \\
2452601.5480  &   $       $& s & $0.0052 $  &$-11017.5$  & $-0.00052  $ & $-0.00074  $& IBVS No. 5378          \\
2452621.3567  &   $       $& p & $0.0010 $  &$-10967  $  & $-0.00001  $ & $-0.00027  $& IBVS No. 5484          \\
2452872.3917  &   $R      $& p & $0.0030 $  &$-10327  $  & $0.00049   $ & $-0.00024  $& IBVS No. 5583          \\
2452872.3917  &   $V      $& p & $0.0029 $  &$-10327  $  & $0.00049   $ & $-0.00024  $& IBVS No. 5583          \\
2452872.3921  &   $I      $& p & $0.0029 $  &$-10327  $  & $0.00089   $ & $0.00016   $& IBVS No. 5583          \\
2452875.5229  &   $V      $& p & $0.0033 $  &$-10319  $  & $-0.00624  $ & $-0.00698  $& IBVS No. 5583          \\
2452875.5298  &   $R      $& p & $0.0035 $  &$-10319  $  & $0.00066   $ & $0.00102   $& IBVS No. 5583          \\
2452875.5309  &   $I      $& p & $0.0034 $  &$-10319  $  & $0.00176   $ & $0.00037   $& IBVS No. 5583          \\
2452878.4734  &   $       $& s & $0.0006 $  &$-10311.5$  & $0.00245   $ & $0.00171   $& IBVS No. 5643          \\
2452896.5148  &   $BVR    $& s & $0.0001 $  &$-10265.5$  & $0.00074   $ & $-0.00003  $& IBVS No. 5668          \\
2452901.4175  &   $BVR    $& p & $0.0001 $  &$-10253  $  & $0.00043   $ & $-0.00036  $& IBVS No. 5668          \\
2452901.6141  &   $BVR    $& s & $0.0002 $  &$-10252.5$  & $0.00091   $ & $0.00012   $& IBVS No. 5668          \\
2452903.3793  &   $BVR    $& p & $0.0001 $  &$-10248  $  & $0.00102   $ & $0.00023   $& IBVS No. 5668          \\
2452925.3448  &   $R      $& p & $0.0001 $  &$-10192  $  & $0.00100   $ & $0.00018   $& IBVS No. 5676          \\
2452925.5409  &   $R      $& s & $0.0001 $  &$-10191.5$  & $0.00098   $ & $0.00016   $& IBVS No. 5676          \\
2452941.4260  &   $       $& p & $0.0003 $  &$-10151  $  & $0.00030   $ & $-0.00055  $& IBVS No. 5643          \\
2452955.3506  &   $I      $& s & $0.0003 $  &$-10115.5$  & $0.00033   $ & $-0.00054  $& IBVS No. 5668          \\
2452956.3321  &   $I      $& p & $0.0003 $  &$-10113  $  & $0.00123   $ & $0.00036   $& IBVS No. 5668          \\
2452957.3119  &   $I      $& s & $0.0001 $  &$-10110.5$  & $0.00043   $ & $-0.00045  $& IBVS No. 5668          \\
2452957.5088  &   $I      $& p & $0.0002 $  &$-10110  $  & $0.00121   $ & $0.00033   $& IBVS No. 5668          \\
2453222.4672  &   $       $& s & $0.001  $  &$-9434.5 $  & $0.00053   $ & $-0.00073  $& IBVS No. 5657          \\
2453233.4496  &   $       $& s & $0.0002 $  &$-9406.5 $  & $0.00017   $ & $-0.00110  $& IBVS No. 5657          \\
2453242.4693  &   $       $& s & $0.0024 $  &$-9383.5 $  & $-0.00168  $ & $-0.00335  $& IBVS No. 5657          \\
2453256.3940  &   $       $& p & $0.0005 $  &$-9348   $  & $-0.00155  $ & $-0.00324  $& IBVS No. 5657          \\
2453256.5921  &   $       $& s & $0.0002 $  &$-9347.5 $  & $0.00043   $ & $-0.00087  $& IBVS No. 5657          \\
2453259.3397  &   $R_c    $& s & $0.0007 $  &$-9340.5 $  & $0.00234   $ & $0.00104   $& IBVS No. 5741          \\
2453259.5344  &   $R_c    $& p & $0.0004 $  &$-9340   $  & $0.00092   $ & $-0.00038  $& IBVS No. 5741          \\
2453259.5359  &   $       $& p & $0.0042 $  &$-9340   $  & $0.00242   $ & $0.00112   $& IBVS No. 5657          \\
2453260.5157  &   $R      $& s & $0.0001 $  &$-9337.5 $  & $0.00162   $ & $0.00031   $& IBVS No. 5676          \\
2453284.4405  &   $       $& s & $0.0065 $  &$-9276.5 $  & $-0.00031  $ & $-0.00202  $& IBVS No. 5657          \\
2453343.282   &   $       $& s & $0.002  $  &$-9126.5 $  & $0.00498   $ & $-0.00164  $& IBVS No. 5653          \\
2453637.4594  &   $       $& s & $0.0003 $  &$-8376.5 $  & $0.00132   $ & $-0.00031  $& IBVS No. 5731          \\
2453704.5331  &   $V      $& s & $0.0003 $  &$-8205.5 $  & $0.00174   $ & $0.00008   $& IBVS No. 5677          \\
2453928.5027  &   $       $& s & $0.0001 $  &$-7634.5 $  & $0.00149   $ & $-0.00019  $& IBVS No. 5777          \\
2453932.4253  &   $       $& s & $0.0008 $  &$-7624.5 $  & $0.00168   $ & $0.00000   $& IBVS No. 5731          \\
2453939.4857  &   $       $& s & $0.0001 $  &$-7606.5 $  & $0.00174   $ & $0.00006   $& IBVS No. 5777          \\
2454004.4019  &   $       $& p & $0.0001 $  &$-7441   $  & $0.00198   $ & $0.00032   $& IBVS No. 5777          \\
2454018.5225  &   $       $& p & $0.0002 $  &$-7405   $  & $0.00189   $ & $0.00023   $& IBVS No. 5777          \\
2454019.5060  &   $       $& s & $0.0004 $  &$-7402.5 $  & $0.00477   $ & $0.00307   $& OEJV No. 0074          \\
2454068.336   &   $       $& p & $0.0003 $  &$-7278   $  & $0.00073   $ & $-0.00090  $& IBVS No. 5777          \\
2454453.3229  &   $       $& s & $0.0003 $  &$-6296.5 $  & $0.00269   $ & $0.00148   $& IBVS No. 5874          \\
2455033.4456  &   $       $& s & $0.0026 $  &$-4817.5 $  & $0.00034   $ & $0.00075   $& IBVS No. 5941          \\
2455051.4876  &   $       $& s & $0.0003 $  &$-4771.5 $  & $-0.00076  $ & $-0.00028  $& IBVS No. 5941          \\
2455062.4707  &   $       $& s & $0.0005 $  &$-4743.5 $  & $-0.00042  $ & $0.00010   $& IBVS No. 5941          \\
2455071.29617 &   $R      $& p & $0.00033$  &$-4721   $  & $-0.00038  $ & $0.00017   $& This paper             \\
2455132.09126 &   $R      $& p & $0.00012$  &$-4566   $  & $-0.00271  $ & $-0.00193  $& This paper             \\
2455141.3114  &   $       $& s & $0.0009 $  &$-4542.5 $  & $-0.00025  $ & $0.00057   $& IBVS No. 5941          \\
2455141.5065  &   $       $& p & $0.0003 $  &$-4542   $  & $-0.00127  $ & $-0.00045  $& IBVS No. 5941          \\
2455445.4929  &   $       $& p & $0.0002 $  &$-3767   $  & $-0.00196  $ & $-0.00015  $& IBVS No. 6010          \\
2455479.4217  &   $       $& s & $0.0007 $  &$-3680.5 $  & $-0.00204  $ & $-0.00016  $& IBVS No. 5984          \\
2455479.4271  &   $       $& s & $0.0007 $  &$-3680.5 $  & $0.00336   $ & $0.00524   $& IBVS No. 5984          \\
2455479.6192  &   $       $& p & $0.0012 $  &$-3680   $  & $-0.00066  $ & $0.00122   $& IBVS No. 5984          \\
2455479.6192  &   $       $& p & $0.0012 $  &$-3680   $  & $-0.00066  $ & $0.00122   $& IBVS No. 5984          \\
2455492.16933 &   $R      $& p & $0.00080$  &$-3648   $  & $-0.00226  $ & $-0.00036  $& This paper             \\
2455501.19018 &   $R      $& p & $0.00025$  &$-3625   $  & $-0.00296  $ & $-0.00104  $& This paper             \\
2455535.11975 &   $R      $& s & $0.00063$  &$-3538.5 $  & $-0.00227  $ & $-0.00030  $& This paper             \\
2455816.15762 &   $R      $& p & $0.00050$  &$-2822   $  & $-0.00537  $ & $-0.00283  $& This paper             \\
2455839.3042  &   $Ir     $& p & $0.0023 $  &$-2763   $  & $-0.00103  $ & $0.00100   $& IBVS No. 6026          \\
2455839.4988  &   $Ir     $& s & $0.0033 $  &$-2762.5 $  & $-0.00255  $ & $-0.00052  $& IBVS No. 6026          \\
2455849.3067  &   $Ir     $& s & $0.0012 $  &$-2737.5 $  & $-0.00069  $ & $0.00133   $& IBVS No. 6026          \\
2455849.5019  &   $Ir     $& p & $0.0019 $  &$-2737   $  & $-0.00161  $ & $0.00041   $& IBVS No. 6026          \\
2455858.7190  &   $       $& s & $0.0004 $  &$-2713.5 $  & $-0.00218  $ & $-0.00017  $& IBVS No. 6011          \\
2455876.17208 &   $R      $& p & $0.00047$  &$-2669   $  & $-0.00385  $ & $-0.00186  $& This paper             \\
2456614.3740  &   $       $& p & $0.0002 $  &$-787    $  & $-0.00026  $ & $0.00025   $& IBVS No. 6167          \\
2456923.06759 &   $VRI    $& p & $0.00015$  &$0       $  & $-0.00066  $ & $-0.00086  $& This paper             \\
2456934.4433  &   $I      $& p & $0.0022 $  &$29      $  & $0.00005   $ & $-0.00018  $& IBVS No. 6152          \\
2456934.6401  &   $I      $& s & $0.0002 $  &$29.5    $  & $0.00073   $ & $0.00050   $& IBVS No. 6152          \\
2456936.2078  &   $VRI    $& s & $0.0011 $  &$33.5    $  & $-0.00056  $ & $-0.00079  $& This paper             \\
2456936.9970  &   $VRI    $& s & $0.0012 $  &$35.5    $  & $0.00415   $ & $0.00391   $& This paper             \\
2456949.3483  &   $I      $& p & $0.0007 $  &$67      $  & $-0.00012  $ & $-0.00039  $& IBVS No. 6152          \\
2456949.5454  &   $I      $& s & $0.0023 $  &$67.5    $  & $0.00085   $ & $0.00059   $& IBVS No. 6152          \\
2457131.5455  &   $I      $& s & $0.0021 $  &$531.5   $  & $0.00094   $ & $0.00029   $& IBVS No. 6196          \\
2457357.2812  &   $UI     $& p & $0.0013 $  &$1107    $  & $0.00171   $ & $0.00061   $& IBVS No. 6196          \\
2457657.3452  &   $       $& p & $0.0002 $  &$1872    $  & $0.00105   $ & $-0.00054  $& OEJV No. 0179          \\
2457657.5416  &   $       $& s & $0.0002 $  &$1872.5  $  & $0.00134   $ & $-0.00025  $& OEJV No. 0179          \\
2458018.405   &   $       $& s & $0.002  $  &$2792.5  $  & $0.00261   $ & $0.00061   $& IBVS No. 6244          \\
\enddata
\label{tab: Minima of v344 Lac}
\end{deluxetable}

\begin{table}
\caption{Obital elements of V344 Lac.}
\label{tab: obital elements}
\footnotesize
\begin{tabular}{lll}
      \hline
Parameter              &   Value                               \\
      \hline
$T_0$[cor]~(HJD)             & 2456923.0683(8)                  \\
$P_0$[cor]~(d)            & 0.39224146(14)                   \\
$\beta$~(d\,yr$^{-1}$)     & $(1.9\pm10.4) \times 10^{-9}$    \\
$A_3$~(d)              & 0.0020(2)                        \\
$a_{12}\sin i^{\prime}$~(au)    & $0.35\pm0.05$                      \\
$e $                     & $0.38\pm0.16$                    \\
$\omega$~(deg)         & $224\pm22$                       \\
$P_3$~(yr)             & $12.4\pm0.5$                     \\
$H_p$~(HJD)            & $2459879.0\pm902.1$              \\
$f(M)~(M_{\odot})$     & $(2.788\pm0.051) \times 10^{-4}$   \\
$\Delta P/P=2{\pi}A_3/P_3$  &$(2.775\pm0.389) \times 10^{-6}$  \\
        \hline
\end{tabular}
\end{table}

\begin{table}
\begin{tiny}
\caption{Photometric solutions for V344 Lac.}
\label{tab: Photometric solutions for V344 Lac}
\begin{tabular}{lcccc}
\hline
Parameters                                &  Photometric elements                 &  uncertainties   &   Photometric elements            &  uncertainties\\
                                          &     without $l_3$                     &                  &   with $l_3$                      &               \\
\hline
$g_1=g_2$                                 &     0.32                              & fixed            &     0.32                          & fixed         \\
$A_1=A_2$                                 &     0.50                              & fixed            &     0.50                          & fixed         \\
$x_{1bolo},x_{2bolo},y_{1bolo},y_{2bolo}$ &     $  0.136, 0.147,0.584,0.572$      & fixed            &     $ 0.136, 0.144,0.584,0.575$   & fixed         \\
$x_{1V},x_{2V},y_{1V},y_{2V}$             &     $  0.139, 0.161,0.673,0.658$      & fixed            &     $ 0.139, 0.155,0.673,0.663$   & fixed         \\
$x_{1R},x_{2R},y_{1R},y_{2R}$             &     $  0.028, 0.046,0.695,0.684$      & fixed            &     $ 0.028, 0.041,0.695,0.687$   & fixed         \\
$x_{1I},x_{2I},y_{1I},y_{2I}$             &     $ -0.038,-0.022,0.665,0.656$      & fixed            &     $-0.038,-0.026,0.665,0.659$   & fixed         \\
$Phase shift$                             &    $ 0.0011$                          & $\pm0.0002$      &    $ 0.0011$                      & $\pm0.0002$   \\
$T_1$\,(K)                                &     6312                              & $\pm54$K         &     6312                          & $\pm54$K      \\
$T_2$                                     &     6166K                             & $\pm64$K         &     6201K                         & $\pm65$K      \\
$q=M_2/M_1$                               &     0.580                             & $\pm0.003$       &     0.387                         & $\pm0.003$    \\
$i$ ($^\circ$)                            &     67.0                              & $\pm0.1$         &     70.3                          & $\pm0.7$      \\
$\Omega_1=\Omega_2$                       &     2.9371                            & $\pm0.0054$      &     2.5457                        & $\pm0.0070$   \\
$\Omega_{in}$                             &     3.0258                            & --               &     2.6512                        & --            \\
$\Omega_{out}$                            &     2.6853                            & --               &     2.4145                        & --            \\
$L_1/(L_1+L_2)(V)$                        &     0.6408                            & $\pm0.0028 $     &     0.7091                        & $\pm0.0221  $ \\
$L_1/(L_1+L_2)(R)$                        &     0.6365                            & $\pm0.0024 $     &     0.7062                        & $\pm0.0214  $ \\
$L_1/(L_1+L_2)(I)$                        &     0.6331                            & $\pm0.0021 $     &     0.7041                        & $\pm0.0210  $ \\
$L_3/(L_1+L_2+L_3)(V)$                    &     --                                & --               &     0.1054                        & $\pm0.0272  $ \\
$L_3/(L_1+L_2+L_3)(R)$                    &     --                                & --               &     0.1014                        & $\pm0.0266  $ \\
$L_3/(L_1+L_2+L_3)(I)$                    &     --                                & --               &     0.0876                        & $\pm0.0265  $ \\
$r_1(pole)$                               &     0.4163                            $\pm0.0010$        &     0.4558                        $\pm0.0015$     \\
$r_1(side)$                               &     0.4441                            $\pm0.0013$        &     0.4925                        $\pm0.0021$     \\
$r_1(back)$                               &     0.4794                            $\pm0.0019$        &     0.5276                        $\pm0.0029$     \\
$r_2(pole)$                               &     0.3263                            $\pm0.0012$        &     0.3018                        $\pm0.0021$     \\
$r_2(side)$                               &     0.3434                            $\pm0.0015$        &     0.3181                        $\pm0.0026$     \\
$r_2(back)$                               &     0.3859                            $\pm0.0027$        &     0.3707                        $\pm0.0057$     \\
$f$                                       &     $26.0\,\%$                        & $\pm1.6\,\%$     &     $44.6\,\%$                    & $\pm3.0\,\%$  \\
$\sum{(O-C)_i^2}$                         &     0.037263                          &                  &     0.037081                      &               \\
\hline
\end{tabular}
\end{tiny}
\end{table}

\begin{table}
\caption{Some contact binaries with cyclic variation of their obital period.}
\label{tab: cyclic variation of contact binaries}
\tiny
\begin{tabular}{llll llll llll ll}
      \hline
Name             &   $P$       &  $q$       &   $M_1$ &  $T_1$ &  $T_2$  & $P_3$    &   $A_3$   & $\Delta P/P$ & $m_3$   &   spot & $l_3$&   $e$        &   reference               \\
                 &      (d)    &($M_2/M_1$) &($M_{\odot}$)  &  (K) &  (K)  & (yr)  &   (d)  & ($\times 10^{-6}$) & ($M_{\odot}$)   &     & (\%)&            &                   \\
      \hline
V776 Cas         &   0.440416842  & 0.130    & 1.55   & 7000 &  6920  &  23.7  &   0.0109  &  7.912   & $1.04   $ &     & 14.9  &  0       &   (1)       \\
V802 Aql         &   0.2677087    & 0.144    & 0.92   & 5000 &  5048  &  8.32  &   0.0196  &  40.526  & $1.52   $ & y   & 0     &  0       &   (2)      \\
MQ UMa           &   0.4760662    & 0.195    &        & 6532 &  6224  &  13.6  &   0.0056  &  7.084   & $       $ &     & 25.1  &  0       &   (3)      \\
VZ Lib           &   0.35825797   & 0.236    & 1.31   & 5446 &  5770  &  2.96  &   0.0039  &  22.666  & $0.52   $ & y   & 10.4  &  0       &   (4)     \\
EZ Hya           &   0.44974752   & 0.257    & 1.37   & 5721 &  6100  &  30.9  &   0.0185  &  10.299  & $0.51   $ &     & 0     &  0       &   (5)       \\
DF CVn           &   0.3269       & 0.285    & 0.93   & 5520 &  4812  &  17.2  &   0.0070  &  7.001   & $0.23   $ & y   & 0     &  0       &   (6)       \\
V1005 Her        &   0.278962083  & 0.301    & 0.92   & 5535 &  5560  &  18.1  &   0.0127  &  12.070  & $0.45   $ & y   & 4.2   &  0       &   (7)       \\
SS Ari           &   0.40598629   & 0.307    & 1.30   & 5488 &  5860  &  37.75 &   0.0112  &  5.104   & $0.278  $ & y   & 0     &  0       &   (8)        \\
GN Boo           &   0.3016022    & 0.320    & 0.99   & 6250 &  6879  &  9.89  &   0.0042  &  7.306   & $0.21   $ &     & 0     &  0       &   (9)      \\
AQ Com           &   0.28133036   & 0.350    & 0.95   & 4911 &  5300  &  8.5   &   0.0016  &  3.238   & $0.08   $ & y   & 0     &  0       &   (10)        \\
GSC 03526-01995  &   0.29225601   & 0.351    & 0.80   & 4581 &  4830  &  7.39  &   0.0090  &  20.858  & $>0.57  $ & y   & 0     &  0       &   (11)      \\
EP And           &   0.4041088    & 0.372    & 1.10   & 6171 &  6250  &  40.89 &   0.0109  &  4.586   & $0.23   $ &     & 9.3   &  0       &   (12)      \\
GSC 1537-1557    &   0.31827564   & 0.378    & 0.95   & 5631 &  5740  &  8.1   &   0.0034  &  7.115   & $>0.19  $ & y   & 0     &  0       &   (13)     \\
RT LMi           &   0.37491782   & 0.378    & 1.29   & 6400 &  6513  &  46.7  &   0.0049  &  1.805   & $       $ &     & 0     &  0       &   (14)      \\
GW Cep           &   0.31883088   & 0.386    & 1.11   & 5800 &  6104  &  32.63 &   0.0089  &  4.692   & $0.22   $ & y   & 0     &  0.076   &   (15)        \\
V344 Lac         &   0.39224141   & 0.387    & 1.16   & 6312 &  6201  &  12.4  &   0.0020  &  2.775   & $0.79  $ &     & 9.81  &  0.38    &   (16)       \\
V396 Mon         &   0.39634132   & 0.392    & 0.92   & 6121 &  6210  &  42.4  &   0.0160  &  6.492   & $0.31   $ & y   & 0.87  &  0       &   (17)       \\
TY UMa           &   0.35454813   & 0.396    & 1.57   & 6229 &  6250  &  51.7  &   0.0182  &  6.056   & $0.434  $ & y   & 0     &  0.73    &   (18)        \\
FV CVn           &   0.31536519   & 0.410    & 0.86   & 5150 &  5465  &  17.65 &   0.0184  &  17.934  & $>0.73  $ &     & 27.7  &  0       &   (19)      \\
PP Lac           &   0.40116138   & 0.435    & 1.18   & 5202 &  5480  &  19.7  &   0.0058  &  5.065   & $0.19   $ &     & 0     &  0       &   (20)      \\
EQ Tau           &   0.3413478    & 0.442    & 1.22   & 5800 &  5740  &  22.7  &   0.0058  &  4.395   & $0.2    $ & y   & 0     &  0.47    &   (21)       \\
UZ CMi           &   0.55136313   & 0.452    &        & 6250 &  6194  &  21.1  &   0.0026  &  2.120   & $       $ &     & $<1$  &  0       &   (22)     \\
LU Lac           &   0.29880186   & 0.480    & 0.95   & 4899 &  5310  &  51.92 &   0.0125  &  4.142   & $0.21   $ &     & 0     &  0       &   (23)      \\
CW Cas           &   0.31886304   & 0.486    & 0.99   & 4950 &  5390  &  69.9  &   0.0320  &  7.875   & $0.52   $ & y   & 9.98  &  0.571   &   (24)      \\
AH Tau           &   0.332672375  & 0.505    & 1.04   & 5840 &  5816  &  54.62 &   0.0349  &  10.992  & $0.767  $ & y   & 0     &  0       &   (25)     \\
V508 Oph         &   0.34479141   & 0.552    & 1.01   & 5980 &  5893  &  24.27 &   0.0036  &  2.552   & $0.1    $ & y   & 0     &  0       &   (26)     \\
V1107 Cas        &   0.27341123   & 0.556    & 0.70   & 4657 &  5200  &  6.74  &   0.0023  &  5.870   & $0.091  $ &     & 1.83  &  0       &   (27)      \\
AL Cas           &   0.5005555    & 0.610    & 1.23   & 6400 &  6136  &  86.6  &   0.0181  &  3.596   & $0.28   $ &     & 0     &  0       &   (28)     \\
PY Lyr           &   0.38576      & 0.660    & 1.51   & 6980 &  7042  &  52.5  &   0.0395  &  12.943  & $1.15   $ & y   & 19.7  &  0.138   &   (29)    \\
PY Vir           &   0.31124849   & 0.773    & 0.95   & 4830 &  4702  &  5.22  &   0.0075  &  24.717  & $0.79   $ &     & 0     &  0       &   (30)      \\
V1044 Her        &   0.24064058   & 0.833    &        & 4377 &  4575  &  14.1  &   0.0026  &  3.172   & $>0.109 $ & y   & 0     &  0.31    &   (31)        \\
FZ Ori           &   0.399986     & 0.886    & 1.17   & 5940 &        &  30.1  &   0.0133  &  7.601   & $0.56   $ &     & 0     &  0       &   (32)     \\
BI Vul           &   0.25182367   & 0.966    &        & 4474 &  4600  &  10.8  &   0.0057  &  9.079   & $       $ & y   & 0     &  0       &   (33)      \\
DM Del           &   0.4198144    &          &        &      &        &  51.22 &   0.0218  &  7.322   & $0.12   $ &     & 0     &  0       &   (34)       \\
V803 Aql         &   0.2634262    &          &        &      &        &  74.6  &   0.0336  &  7.748   & $0.52   $ &     & 0     &  0       &   (29)    \\
      \hline
\end{tabular}
Reference: (1) \cite{Zhouetal2016}; (2) \cite{Yangetal2008}; (3) \cite{Zhouetal2015}; (4) \cite{Liaoetal2019}; (5) \cite{Yangetal2004}; (6) \cite{Daietal2011}; (7) \cite{Zhuetal2019}; (8) \cite{Liuetal2009}; (9) \cite{Yangetal2013}; (10) \cite{Liuetal2014}; (11) \cite{Liaoetal2012}; (12) \cite{Liaoetal2013}; (13) \cite{Xiangetal2015a}; (14) \cite{Qianetal2008}; (15) \cite{Leeetal2010}; (16) The present paper; (17) \cite{Liuetal2011}; (18) \cite{Lietal2015}; (19) \cite{LiaoSarotsakulchai2019}; (20) \cite{Qianetal2005}; (21) \cite{Lietal2014}; (22) \cite{Qianetal2013a}; (23) \cite{Liaoetal2014}; (24) \cite{Wangetal2014}; (25) \cite{Xiangetal2015b}; (26) \cite{Xiangetal2015c}; (27) \cite{Liuetal2016}; (28) \cite{Qianetal2014a}; (29) \cite{Zascheetal2009}; (30) \cite{Zhuetal2013}; (31) \cite{Luetal2016}; (32) \cite{Prasadetal2014}; (33) \cite{Qianetal2013b}; (34) \cite{HeQian2010}.
\end{table}

\begin{thebibliography}{}
\bibitem[\protect\citeauthoryear{Agerer \& Hubscher}{2003}]{AgererHubscher2003}Agerer, F., Hubscher, J., 2003, IBVS No. 5484, 1
\bibitem[\protect\citeauthoryear{Applegate}{1992}]{Applegate1992}Applegate J. H. 1992, ApJ, 385, 621
\bibitem[\protect\citeauthoryear{Binnendijk}{1970}]{Binnendijk1970}Binnendijk, L. 1970, Vistas Astron., 12, 217
\bibitem[\protect\citeauthoryear{Brat et al.}{2007}]{Bratetal2007}Brat, L., Zejda, M., Svoboda, P. 2007, OEJV No. 74, 1
\bibitem[\protect\citeauthoryear{Claret \& Gimenez}{1990}]{ClaretGimenez1990}Claret, A., and Gimenez, A., 1990, A\&A 230, 412
\bibitem[\protect\citeauthoryear{Cox}{2000}]{Cox2000}Cox, A. N. 2000, Allen' s Astrophysical Quantities (4th ed., NewYork: Springer)
\bibitem[\protect\citeauthoryear{Dai et al.}{2011}]{Daietal2011}Dai, H.-F., Yang, Y.-G., \& Yin, X.-G. 2011, NewA, 16, 173
\bibitem[\protect\citeauthoryear{D'Angelo et al.}{2006}]{D'Angeloetal2006}D'Angelo, C., van Kerkwijk, M. H., Rucinski, S. M. 2006, AJ, 132, 650
\bibitem[\protect\citeauthoryear{Diethelm}{2003}]{Diethelm2003}Diethelm, R. 2003 IBVS No. 5438, 1
\bibitem[\protect\citeauthoryear{Diethelm}{2005}]{Diethelm2005}Diethelm, R. 2005, IBVS No. 5653, 1
\bibitem[\protect\citeauthoryear{Diethelm}{2012}]{Diethelm2012}Diethelm, R. 2012, IBVS No. 6011, 1
\bibitem[\protect\citeauthoryear{Dvorak}{2003}]{Dvorak2003}Dvorak, S. W. 2003, IBVS No. 5378, 1
\bibitem[\protect\citeauthoryear{Dvorak}{2006}]{Dvorak2006}Dvorak, S. W. 2006, IBVS No. 5677, 1
\bibitem[\protect\citeauthoryear{Eggen}{1967}]{Eggen1967}Eggen, O. J. 1967, MemRAS, 70, 111
\bibitem[\protect\citeauthoryear{Eggleton}{2012}]{Eggleton2012}Eggleton, P. P. 2012, JASS, 29, 145
\bibitem[\protect\citeauthoryear{Fabrycky \& Tremaine}{2007}]{FabryckyTremaine2007}Fabrycky, D. \& Tremaine, S. 2007, AJ, 669, 1298
\bibitem[\protect\citeauthoryear{Fan et al.}{2015}]{Fanetal2015}Fan Y. F., Bai J. M., Zhang J. J., Wang C. J., Chang L., Xin Y. X., Zhang R. L., 2015, Res. Astron. Astrophys., 15, 918
\bibitem[\protect\citeauthoryear{Flannery}{1976}]{Flannery1976}Flannery, B. P. 1976, ApJ, 205, 217
\bibitem[\protect\citeauthoryear{Gaia Collaboration et al.}{2018}]{GaiaCollaborationetal2018}Gaia Collaboration, Brown, A. G. A., Vallenari, A et al. 2018, A\&A, 616A, 1
\bibitem[\protect\citeauthoryear{He \& Qian}{2010}]{HeQian2010}He, J.-J. \& Qian, S.-B. 2010, PASJ, 62, 441
\bibitem[\protect\citeauthoryear{Hoffmann}{1983}]{Hoffmann1983}Hoffmann, M. 1983, IBVS No. 2344, 1
\bibitem[\protect\citeauthoryear{Hu et al.}{2014}]{Huetal2014}Hu Shao-Ming, Han Sheng-Hao, Guo Di-Fu and Du Jun-Ju 2014, RAA, Vol. 14 No. 6, 719¨C732
\bibitem[\protect\citeauthoryear{Hubscher \& Lehmann}{2012}]{HubscherLehmann2012}Hubscher, J., Lehmann, P. B. 2012, IBVS No. 6026, 1
\bibitem[\protect\citeauthoryear{Hubscher et al.}{2005}]{Hubscheretal2005}Hubscher, J., Paschke, A., Walter, F. 2005, IBVS No. 5657, 1
\bibitem[\protect\citeauthoryear{Hubscher et al.}{2006}]{Hubscheretal2006}Hubscher, J., Paschke, A., Walter, F. 2006, IBVS No. 5731, 1
\bibitem[\protect\citeauthoryear{Hubscher et al.}{2009}]{Hubscheretal2009}Hubscher, J., Steinbach, H.-M., Walter, F. 2009, IBVS No. 5874, 1
\bibitem[\protect\citeauthoryear{Hubscher et al.}{2010}]{Hubscheretal2010}Hubscher, J., Lehmann, P. B., Monninger, G., Steinbach, H.-M., Walter, F. 2010, IBVS No. 5941, 1
\bibitem[\protect\citeauthoryear{Hubscher et al.}{2012}]{Hubscheretal2012}Hubscher, J., Lehmann, P. B., Walter, F. 2012, IBVS No. 6010, 1
\bibitem[\protect\citeauthoryear{Hubscher}{2005}]{Hubscher2005}Hubscher, J., 2005, IBVS No. 5643, 1
\bibitem[\protect\citeauthoryear{Hubscher}{2011}]{Hubscher2011}Hubscher, J. 2011, IBVS No. 5984, 1
\bibitem[\protect\citeauthoryear{Hubscher}{2015}]{Hubscher2015}Hubscher, J. 2015, IBVS No. 6152, 1
\bibitem[\protect\citeauthoryear{Irwin}{1952}]{Irwin1952}Irwin, J. B. 1952, ApJ, 116, 211
\bibitem[\protect\citeauthoryear{Irwin}{1959}]{Irwin1959}Irwin, J. B. 1959, ApJ, 64, 149
\bibitem[\protect\citeauthoryear{Kholopov et al.}{1985}]{Kholopovetal1985}Kholopov, P. N., Samus, N. N., Kazarovets, E. V., Perova, N. B. 1985, IBVS No. 2681, 1
\bibitem[\protect\citeauthoryear{Koleva et al.}{2009}]{Kolevaetal2009}Koleva, M., Prugniel, Ph., Bouchard, A., Wu, Y. 2009, A\&A, 501, 1269
\bibitem[\protect\citeauthoryear{Kopal}{1978}]{Kopal1978}Kopal, D.: Dynamics of Close Binary Systems. D. Reidel, Dordrecht(1978)
\bibitem[\protect\citeauthoryear{Kotkova \& Wolf}{2006}]{KotkovaWolf2006}Kotkova, L., Wolf, M. 2006, IBVS No. 5676, 1
\bibitem[\protect\citeauthoryear{Kozai}{1962}]{Kozai1962}Kozai, Y. 1962, AJ, 67, 591
\bibitem[\protect\citeauthoryear{Lanza et al.}{1998}]{Lanzaetal1998}Lanza A. F., Rodono M. Rosner R., 1998, MNRAS, 296, 893
\bibitem[\protect\citeauthoryear{Lee et al.}{2010}]{Leeetal2010}Lee, J. W., Youn, J.-H., Han, W.-Y., et al. 2010, AJ, 139, 898
\bibitem[\protect\citeauthoryear{Li et al.}{2014}]{Lietal2014}Li, K., Qian, S. -B., Hu, S. -M. \& He, J. -J.
\bibitem[\protect\citeauthoryear{Li et al.}{2015}]{Lietal2015}Li, K., Hu, S. -M., Guo, D. -F., Jiang, Y. -G., Gao, Y. -G., Chen, X. \& Odell, Andrew P 2015, AJ, 149, 120L
\bibitem[\protect\citeauthoryear{Liao \& Qian}{2010}]{LiaoQian2010}Liao, W.-P. \& Qian, S.-B. 2010, MNRAS, 405, 1930
\bibitem[\protect\citeauthoryear{Liao et al.}{2012}]{Liaoetal2012}Liao, W. -P., Qian, S. -B. \& Liu, N. -P. 2012, AJ, 144, 178L
\bibitem[\protect\citeauthoryear{Liao et al.}{2013}]{Liaoetal2013}Liao, W. -P., Qian, S. -B., Li, K., He, J. -J., Zhao, E. -G. \& Zhou, X. 2013, AJ, 146, 79L
\bibitem[\protect\citeauthoryear{Liao et al.}{2014}]{Liaoetal2014}Liao, W. -P., Qian, S. -B., Zhao, E. -G. \& Jiang, L. -Q. 2014, NewA, 31, 65L
\bibitem[\protect\citeauthoryear{Liao et al.}{2019}]{Liaoetal2019}Liao, W. -P., Qian, S. -B. \& Sarotsakulchai, T. 2019, AJ, 157, 207L
\bibitem[\protect\citeauthoryear{Liao \& Sarotsakulchai}{2019}]{LiaoSarotsakulchai2019}Liao, W. -P. \& Sarotsakulchai, T. 2019, PASP, 131a, 4202L
\bibitem[\protect\citeauthoryear{Liu et al.}{2007}]{Liuetal2007}Liu, Liang, Qian, Sheng-Bang, Boonrucksar, Soonthornthum, Zhu, Li-Ying, He, Jia-Jia and Yuan, J. -Z. 2007, PASJ, 59, 607
\bibitem[\protect\citeauthoryear{Liu et al.}{2009}]{Liuetal2009}Liu, L., Qian, S. B., He, J.-J., Zhang, J. \& Li, L.-J. 2009, Ap\&SS, 321, 19
\bibitem[\protect\citeauthoryear{Liu et al.}{2011}]{Liuetal2011}Liu, L., Qian, S. -B., Liao, W. -P., He, J. -J., Zhu, L. -Y., Li, L. -J. \& Zhao, E. -G. 2011, AJ, 141, 44L
\bibitem[\protect\citeauthoryear{Liu et al.}{2016}]{Liuetal2016}Liu, Liang, Qian, Shengbang, He, Jiajia, Liao, Wenping \& Liu, Nianping 2016, PASJ, 68, 31L
\bibitem[\protect\citeauthoryear{Liu et al.}{2018}]{Liuetal2018}Liu, L., Qian, S. B. \& Xiong, X. 2018, MNRAS, 474, 5199
\bibitem[\protect\citeauthoryear{Liu et al.}{2014}]{Liuetal2014}Liu, N.-P., Qian, S.-B., Wang, J.-J., et al. 2014, NewA, 32, 31
\bibitem[\protect\citeauthoryear{Lu et al.}{2016}]{Luetal2016}Lu, Hongpeng, Zhang, Liyun, Han, Xianming L., Pi, Qingfeng \& Wang, Daimei 2016, NewA, 48, 58L
\bibitem[\protect\citeauthoryear{Lucy}{1967}]{Lucy1967}Lucy, L. B. 1967, Zeitschrift f$\rm{\ddot{u}}$r Astrophysik, 65, 89
\bibitem[\protect\citeauthoryear{Lucy}{1976}]{Lucy1976}Lucy, L. B. 1976, ApJ, 205, 208
\bibitem[\protect\citeauthoryear{Miller \& Wachmann}{1973}]{MillerWachmann1973}Miller, W. J. \& Wachmann, A. A. 1973, RA, 8, 367
\bibitem[\protect\citeauthoryear{Molik \& Wolf}{2004}]{MolikWolf2004}Molik P. \& Wolf, M. 2004, BaltA, 13, 145
\bibitem[\protect\citeauthoryear{Parimucha et al.}{2007}]{Parimuchaetal2007}Parimucha, S., Vanko, M., Pribulla, T. et al. 2007, IBVS No. 5777, 1
\bibitem[\protect\citeauthoryear{Parimucha et al.}{2016}]{Parimuchaetal2016}Parimucha, S., Dubovsky, P., Kudak, V., Perig, V. 2016, IBVS No. 6167, 1
\bibitem[\protect\citeauthoryear{Prasad et al.}{2014}]{Prasadetal2014}Prasad, Vinod, Pandey, J. C., Patel, M. K. \& Srivastava, D. C. 2014, Ap\&SS, 353, 575P
\bibitem[\protect\citeauthoryear{Pribulla \& Rucinski}{2006}]{PribullaRucinski2006}Pribulla, T. \& Rucinski, S. M. 2006, AJ, 131, 2986
\bibitem[\protect\citeauthoryear{Pribulla et al.}{2005a}]{Pribullaetal2005a}Pribulla, T., Vanko, M., Chochol, D., Parimucha, S., Baludansky, D. 2005, Ap\&SS, 296, 281
\bibitem[\protect\citeauthoryear{Pribulla et al.}{2005b}]{Pribullaetal2005b}Pribulla, T., Baludansky, D., Chochol, D. et al. 2005, IBVS No. 5668, 1
\bibitem[\protect\citeauthoryear{Qian et al.}{2005}]{Qianetal2005}Qian, S.-B., Zhu, L.-Y., Soonthornthum, B. 2005, NewA, 11, 52
\bibitem[\protect\citeauthoryear{Qian et al.}{2008}]{Qianetal2008}Qian, S.-B., He, J.-J. \& Xiang, F.-Y. 2008, PASJ, 60, 77
\bibitem[\protect\citeauthoryear{Qian et al.}{2013a}]{Qianetal2013a}Qian, S. -B., Li, K., Liao, W. -P., Liu, L., Zhu, L. -Y., He, J. -J., Wang, J. -J. \& Zhao, E. -G. 2013a, AJ, 145, 91Q
\bibitem[\protect\citeauthoryear{Qian et al.}{2013b}]{Qianetal2013b}Qian, S.-B., Liu, N.-P., Li, K., et al. 2013b, ApJS, 209, 13
\bibitem[\protect\citeauthoryear{Qian et al.}{2013c}]{Qianetal2013c}Qian, S.-B., Zhang, J., Wang, J.-J., et al. 2013c, ApJS, 207, 22
\bibitem[\protect\citeauthoryear{Qian et al.}{2014b}]{Qianetal2014b}Qian, S.-B., Wang, J.-J., Zhu, L.-Y., et al. 2014b, ApJS, 212, 4
\bibitem[\protect\citeauthoryear{Qian et al.}{2014a}]{Qianetal2014a}Qian, S. -B., Zhou, X., Zola, S., Zhu, L. -Y., Zhao, E. -G., Liao, W. -P. \& Leung, K. -C. 2014a, AJ, 148, 79Q
\bibitem[\protect\citeauthoryear{Qian et al.}{2017}]{Qianetal2017}Qian, S.-B., He, J.-J., Zhang, J., et al. 2017, RAA, 17, 87
\bibitem[\protect\citeauthoryear{Qian et al.}{2018}]{Qianetal2018}Qian, S.-B., Zhang, J., He, J.-J., Zhu, L.-Y., Zhao, E.-G., Shi, X.-D., Zhou, X., Han, Z.-T. 2018, ApJS, 235, 5
\bibitem[\protect\citeauthoryear{Robertson \& Eggleton}{1977}]{RobertsonEggleton1977}Robertson, J. A. \& Eggleton, P. P. 1977, MNRAS, 179, 359
\bibitem[\protect\citeauthoryear{Rucinski et al.}{2007}]{Rucinskietal2007}Rucinski, S. M., Pribulla, T., van K. M. H. 2007, AJ, 134, 2353
\bibitem[\protect\citeauthoryear{Rucinski}{1969}]{Rucinski1969}Rucinski, S. M. 1969, AcA, 19, 245
\bibitem[\protect\citeauthoryear{Safar \& Zejda}{2000}]{SafarZejda2000}Safar, J., Zejda, M. 2000, IBVS No. 4887, 1
\bibitem[\protect\citeauthoryear{Safar \& Zejda}{2002}]{SafarZejda2002}Safar, J., Zejda, M. 2002, IBVS No. 5263, 1
\bibitem[\protect\citeauthoryear{Tokovinin et al.}{2006}]{Tokovininetal2006}Tokovinin, A., Thomas, S., Sterzik, M., Udry, S. 2006, A\&A, 450, 681
\bibitem[\protect\citeauthoryear{Tout \& Hall}{1991}]{Touthall1991}Tout, C. A., Hall, D. S. 1991, MNRAS, 253, 9T
\bibitem[\protect\citeauthoryear{van Hamme \& Wilson}{2007}]{vanHammeWilson2007}Van Hamme, W., \& Wilson, R. E. 2007, ApJ, 661, 1129
\bibitem[\protect\citeauthoryear{Wang et al.}{2014}]{Wangetal2014}Wang, J.-J., Qian, S.-B., He, J.-J., Li, L. J. \& Zhao, E. G. 2014, AJ, 148, 95
\bibitem[\protect\citeauthoryear{Williams et al.}{2015}]{Williamsetal2015}Williams, S. C., Darnley, M. J., Bode, M. F., Steele, I. A. 2015, ApJ, 805L, 18
\bibitem[\protect\citeauthoryear{Wilson \& Devinney}{1971}]{WilsonDevinney1971}Wilson, R. E., \& Devinney, E. J. 1971, ApJ, 166, 605
\bibitem[\protect\citeauthoryear{Wilson \& van Hamme}{2014}]{WilsonvanHamme2014}Wilson, R. E., \& van Hamme, W. 2014, ApJ, 780, 151
\bibitem[\protect\citeauthoryear{Wilson et al.}{2010}]{Wilsonetal2010}Wilson, R. E., Van Hamme, W., \& Terrell, D. 2010, ApJ, 723, 1469
\bibitem[\protect\citeauthoryear{Wilson}{1979}]{Wilson1979}Wilson, R. E. 1979, ApJ, 234, 1054
\bibitem[\protect\citeauthoryear{Wilson}{1990}]{Wilson1990}Wilson, R. E. 1990, ApJ, 356, 613
\bibitem[\protect\citeauthoryear{Wilson}{2008}]{Wilson2008}Wilson, R. E. 2008, ApJ, 672, 575
\bibitem[\protect\citeauthoryear{Wilson}{2012}]{Wilson2012}Wilson, R. E. 2012, AJ, 144, 73
\bibitem[\protect\citeauthoryear{Worthey \& Lee}{2011}]{WortheyLee2011}Worthey, G., \& Lee, H.-C. 2011, ApJS, 193, 1
\bibitem[\protect\citeauthoryear{Wu et al.}{2011}]{Wuetal2011}Wu, Yue, Singh, H. P., Prugniel, P., Gupta, R., Koleva, M. 2011, A\&A, 525A, 71
\bibitem[\protect\citeauthoryear{Xiang et al.}{2015a}]{Xiangetal2015a}Xiang, F. -Y., Xiao, T. -Y., Zhang, B. \& Shi, X. -D. 2015a, AJ, 150, 9X
\bibitem[\protect\citeauthoryear{Xiang et al.}{2015b}]{Xiangetal2015b}Xiang, Fu-Yuan, Xiao, Ting-Yu \& Yu, Yun-Xia 2015b, AJ, 150, 25X
\bibitem[\protect\citeauthoryear{Xiang et al.}{2015c}]{Xiangetal2015c}Xiang, Fu-Yuan, Yu, Yun-Xia \& Xiao, Ting-Yu 2015c, AJ, 149, 62X
\bibitem[\protect\citeauthoryear{Yang et al.}{2004}]{Yangetal2004}Yang, Y.-G., Qian, S.-B. \& Zhu, C.-H. 2004, PASP, 116, 826
\bibitem[\protect\citeauthoryear{Yang et al.}{2008}]{Yangetal2008}Yang, Y.-G., Qian, S.-B., Zhu, L.-Y., et al. 2008, PASJ, 60, 803
\bibitem[\protect\citeauthoryear{Yang et al.}{2009}]{Yangetal2009}Yang, Y.-G., Qian, S.-B., Zhu, L.-Y. and He, J.-J. 2009, AJ, 138, 540
\bibitem[\protect\citeauthoryear{Yang et al.}{2013}]{Yangetal2013}Yang, Y.-G., Li, H.-L. \& Dai, H.-F. 2013, AJ, 145, 60
\bibitem[\protect\citeauthoryear{Zacharias et al.}{2015}]{Zachariasetal2015}Zacharias, N., Finch, C., Subasavage, J., et al. 2015, AJ, 150, 101
\bibitem[\protect\citeauthoryear{Zasche et al.}{2009}]{Zascheetal2009}Zasche, P., Liakos, A., Niarchos, P., et al. 2009, NewA, 14, 121
\bibitem[\protect\citeauthoryear{Zejda et al.}{2006}]{Zejdaetal2006}Zejda, M., Mikulasek, Z., Wolf, M. 2006, IBVS No. 5741, 1
\bibitem[\protect\citeauthoryear{Zejda}{2004}]{Zejda2004}Zejda, M. 2004, IBVS No. 5583, 1
\bibitem[\protect\citeauthoryear{Zhang et al.}{2009}]{Zhangetal2009}Zhang, X.-B., Deng, L. and  Lu, P. 2009, AJ, 138, 680
\bibitem[\protect\citeauthoryear{Zhou et al.}{2009}]{Zhouetal2009}Zhou A.-Y., Jiang X.-J., Zhang Y.-P., Wei J.-Y. 2009, RAA, 9, 349
\bibitem[\protect\citeauthoryear{Zhou et al.}{2015}]{Zhouetal2015}Zhou, X., Qian, S.-B., Liao, W.-P., Zhao, E.-G., Wang, J.-J. and Jiang, L.-Q. 2015, AJ, 150, 83
\bibitem[\protect\citeauthoryear{Zhou et al.}{2016}]{Zhouetal2016}Zhou, X., Qian, S.-B., Zhang, J., Jiang, L.-Q., Zhang, B. and Kriener, J. 2016, ApJ, 817, 133
\bibitem[\protect\citeauthoryear{Zhu et al.}{2013}]{Zhuetal2013}Zhu, L. Y., Qian, S. B., Liu, N. P., Liu, L. \& Jiang, L. Q.
\bibitem[\protect\citeauthoryear{Zhu et al.}{2019}]{Zhuetal2019}Zhu, L.-Y., Wang, Z.-H., Tian, X.-M., Li, L.-J. and Gao, X. 2019, MNRAS, 489, 2677
\end{thebibliography}
\end{document}